\documentstyle[aps,prb]{revtex}
\begin{document}
\draft
\title{Electronic Raman scattering in Tl$_2$Ba$_2$CuO$_{6+\delta}$:
symmetry of the order parameter, oxygen doping effects, and normal
state scattering}

\author{L.V. Gasparov}
\address{2. Physikalisches Institut, RWTH-Aachen, 52056 Aachen, Germany
and Institute for Solid State Physics, 142432, Chernogolovka, Moscow
district,
Russia}

\author{P. Lemmens}
\address{2. Physikalisches Institut, RWTH-Aachen, 52056 Aachen, Germany}

\author{N.N. Kolesnikov}
\address{Institute for Solid State Physics 142432, Chernogolovka,
Moscow district, Russia}

\author{G. G\"untherodt}
\address{2. Physikalisches Institut, RWTH-Aachen, 52056 Aachen, Germany}
\date{\today}
\maketitle
\begin{abstract}
Single crystals of the optimally doped, moderately  and strongly
overdoped high temperature superconductor
Tl$_2$Ba$_2$CuO$_{6+\delta}$ (Tl-2201) with T$_c$=80, 56 and 30~K,
respectively, have been investigated by polarized Raman
scattering. By taking the peak position of the B$_{1g}$ component
of  electronic Raman scattering as 2$\Delta_0$ we found that the
reduced gap value (2$\Delta_0/k_BT_c$) strongly decreases with
increasing doping. The behavior of the low frequency scattering
for the B$_{1g}$ and B$_{2g}$ scattering components is similar for
optimally doped and overdoped crystals and can be described by a
$\omega^3$- and $\omega$ -law, respectively, which is consistent
with a d-wave symmetry of the order parameter. In contrast to the
optimally doped Tl-2201 in both, moderately and strongly overdoped
Tl-2201, the relative (compared to the B$_{1g}$) intensity of the
A$_{1g}$ scattering component is suppressed. We suggest that the
van Hove singularity is responsible for the observed changes of
Raman intensity and reduced gap value with doping. Electronic
Raman scattering in the normal state is discussed in the context
of the scattering from impurities and compared to the existing
infrared data. The scattering rate evaluated from the Raman
measurements is smaller for the overdoped samples, compared to the
moderately overdoped samples.

\end{abstract}

\section*{Introduction}

The symmetry of the order parameter is one of the most important questions
for
the high temperature superconductors (HTSC).  This issue is especially
interesting as a function of different doping levels.  Electronic Raman
scattering
(ELRS) plays a special role in addressing this problem
\cite{Abr73,Kle84,Dev95,Dev97,Car96,Carb96,Weng97}. The symmetry properties
of
the order parameter can be determined by investigating the anisotropy of
the scattering cross section for the different symmetry components.
Different scattering components originate from different areas of the
Fermi surface (FS). The ratio of one scattering component compared with
another one reflects the changes of the Fermi surface (FS) topology with
doping\cite{Chen97}. There are several theoretical attempts
\cite{Dev95,Dev97,Car96,Carb96,Weng97}
to describe the electronic Raman scattering  in HTSC at $T<T_c$,
but still there is no consensus concerning the exact  mechanism of the
scattering.

In the optimally doped HTSC the electronic Raman scattering from single
crystals in the superconducting state reveals several common features
\cite
{Dev95,Dev97,Car96,Carb96,Weng97,Hackl88,Coop88,Chen93,Stauf92,Hof94,Chen94,Nem93,Gasp97}.
The superconducting transition manifests itself in a redistribution of the
ELRS continuum  into a broad peak (pair-breaking peak), the
intensity and frequency position $\Omega$ of which  differs for
the different symmetry components. For the optimally doped samples, one has
$\Omega(B_{1g}) > \Omega(B_{2g}) > \Omega(A_{1g})$
\cite
{Dev95,Dev97,Car96,Carb96,Weng97,Hackl88,Coop88,Chen93,Stauf92,Hof94,Chen94,Nem93,Gasp97}.
The scattering on the low frequency side of the pair-breaking peak
does not reveal additional peaks or a cut-off, which would be an
indication of anisotropic s-wave component. In contrast, a
power-law decrease of the scattering intensity toward zero frequency shift
is observed. In the B$_{1g}$ scattering component this power-law is close
to a $\omega^3$-dependence, while in the A$_{1g}$  and B$_{2g}$ scattering
components a linear-in-$\omega$ decrease is observed. The above mentioned
features were first described by Devereaux et al.\cite{Dev95} in the
framework of a d-wave order parameter, i.e. using the gap function
$\Delta(\vec{k})=\Delta_{max}\cos2\phi$, where $\phi$ is an angle between
$\vec{k}$ and Cu-O bond direction within the CuO$_2$ plane.

The general description of the Raman scattering cross section follows
from the fluctuation-dissipation theorem. For the case of nonresonant
scattering the Raman scattering cross section is given by the
Raman response function $\chi_{\gamma,\gamma}(\vec{q},\omega:)$

\begin{equation}
\frac{\partial^2\sigma}{\partial\omega\partial\Omega}\propto
\left[1+n(\omega)\right]\Im
\mbox{m}\,\chi_{\gamma\gamma}(\vec{q},\omega),
\end{equation}

where $n(\omega)=1/(\exp(\omega/T)-1)$ is the  Bose-factor,
$\omega$=$\omega_I-\omega_S$, is Stokes Raman shift,
where  $\omega_I(\omega_S)$ is the
frequency of the incident (scattered) photon.

The Raman response function due to the breaking of Cooper pairs in
a superconductor  and including Coulomb repulsion can be written
as \cite{Abr73,Kle84}:

\begin{equation}\label{rares}
\chi_{\gamma\gamma}(\vec{q},\omega)=\langle\gamma^2_{\vec{k}}
\lambda_{\vec{k}}\rangle
-\frac{\langle\gamma_{\vec{k}}\lambda_{\vec{k}}\rangle^2}
{\langle\lambda_{\vec{k}}\rangle},
\end{equation}

where  $\gamma_{\vec{k}}$ is the  Raman vertex, which describes the strength
of the corresponding Raman transition, $\lambda_{\vec{k}}$ is the Tsuneto
function\cite{Tsun60} and the brackets $\langle\cdots\rangle$ denote  the
average of the momentum $\vec{k}$ over the Fermi surface.

The Tsuneto function is determined as:

\begin{equation}\label{Tsuneto}
\lambda({\vec{k},i\omega})=
\frac{\Delta(\vec{k})^2}{E(\vec{k})}
\tanh{\left[\frac{E(\vec{k})}{2T}\right]}
\left[\frac{1}{2E(\vec{k})+i\omega}
+\frac{1}{2E(\vec{k})-i\omega}\right],
\end{equation}
with excitation energy E$^2(\vec{k})=\xi^2(\vec{k})+\Delta^2(\vec{k})$,
conduction band $\xi(\vec{k})=\epsilon(\vec{k})-\mu$, chemical potential
$\mu$ and superconducting energy gap $\Delta(\vec{k})$.

There is an important consequence following from Eqs. \ref{rares}
and \ref{Tsuneto} that the Raman response is proportional to the
square of the superconducting order parameter, therefore, as it
was already mentioned in Ref.\onlinecite{Car96}, Raman scattering
is not sensitive to the sign of the order parameter.

In the case of nonresonant scattering one can describe the Raman
vertex through the curvature of the energy bands
$\epsilon(\vec{k})$:

\begin{equation}
\gamma_{\vec{k}}=\sum_{\alpha \beta}e_\alpha^I
\frac{\partial^2\epsilon(\vec{k})}{\partial k_{\alpha}\partial k_{\beta}}
e^S_{\beta},
\end{equation}
where $\vec{e}^I(\vec{e}^S)$ is  the polarization vector of the incident
(scattered) photon, and $\alpha$ and $\beta$ are summation indices
which correspond to the different projections of $\vec{k}$.

If one assumes\cite{Dev95} that the Raman vertex does not depend on the
frequency of the incident photon one can take into account symmetry
considerations to evaluate corresponding Raman scattering
components. In such a case the Raman vertex can be described
in terms of Brillouin zone (BZ) or Fermi surface harmonics\cite{Dev95}
$\Phi_L(\vec{k})$, which transform according to point group transformations
of the crystal.

\begin{equation}
\gamma_{\vec{k}}(\omega_i,\omega_s)=\sum_L\gamma_L(\omega_i,\omega_s)
\Phi_L(\vec{k}).
\end{equation}

For tetragonal symmetry one gets the following form of the
Raman vertices\cite{Dev95}:

\begin{equation}
\gamma_{B_{1g}} \propto \cos2\phi\qquad
\gamma_{B_{2g}} \propto \sin2\phi\qquad
\gamma_{A_{1g}} \propto 1+\cos4\phi.
\end{equation}

Let us analyze the Raman response (Eq.\ref{rares}). For simplicity we have
drawn
in Fig.1 corresponding polar plots of the functions contained in each of the
two
terms of the Raman response. The first term is the "bare" Raman response
which
reflects the attractive interaction in the Cooper pair whereas the second
term
("screening") is due to the Coulomb repulsion. Let us start with the
"screening"term. This term is proportional to the squared FS average of the
product of the Raman vertex $\gamma$ and the Tsuneto function $\lambda$. The
Tsuneto function in turn is proportional to the square of the gap function.
Following Devereaux and Einzel\cite{Dev95} we assume a d-wave gap in the
form of
$\Delta(\vec{k})=\Delta_{max}\cos2\phi$, which has a B$_{1g}$ symmetry. When
squared it becomes totally symmetric (A$_{1g}$). Therefore an averaged
product
of the Raman vertex and Tsuneto function will be nonzero only if the vertex
function is totally symmetric. This is not the case for the B$_{1g}$
($\gamma\sim\cos2\phi$) and B$_{2g}$ ($\gamma\sim\sin2\phi$) Raman vertexes,
but
only for the A$_{1g}$ ($\gamma\sim1+\cos4\phi$) as seen in Fig.1. Therefore
A$_{1g}$ scattering component is the only component strongly affected or
"screened" by the long range Coulomb interaction
\cite{Dev95,Dev97,Car96,Carb96,Weng97}. Let us now look on the bare Raman
response. This term is proportional to the FS average of the product of the
squared Raman vertex $\gamma^2$ and Tsuneto function $\lambda$
($\propto\Delta(\vec{k})^2$). Both $\gamma^2$ and $\lambda$ are totally
symmetric. One sees from Fig.1 that maxima and nodes of the squared B$_{1g}$
Raman vertex coincide with that of squared d-wave gap. This leads to the
highest
relative peak position for the B$_{1g}$ scattering component and a
$\omega^3$-dependence of the low frequency scattering. In contrast, maxima
of
the B$_{2g}$ Raman vertex coincide with nodes of the squared d-wave order
parameter, resulting in a lower relative peak position and a
linear-in-$\omega$
low frequency dependence for this component. The A$_{1g}$ scattering
component
is the only one which is screened. The "screening" term shifts the peak
position
of the A$_{1g}$ scattering component to a frequency smaller than that of the
B$_{1g}$. Because of the "screening" term, one could expect that the
A$_{1g}$
ELRS peak should be the weakest one\cite{Car96,Carb96,Weng97}. Nevertheless
in
all optimally doped HTSC (YBCO\cite{Hackl88,Coop88,Chen93},
Bi-2212\cite{Stauf92}, Tl-2223\cite{Hof94}, La-214\cite{Chen94},
Tl-2201\cite{Nem93,Gasp97}) the relative intensity of the A$_{1g}$ ELRS peak
is
strong and comparable to that of the B$_{1g}$ peak. This contradicts
existing
LDA-based calculations of the electronic Raman scattering
cross-section\cite{Car96}. However, resonance effects\cite{Blum96,Sher97}
may
alter these calculations. This picture qualitatively describes the
experimental
results for all optimally doped HTSC's. The only exception is the $n$-type
superconductor (Nd,Ce)-214, which demonstrates a behavior consistent with
an s-wave type of order parameter\cite{Hacl95}.

For the overdoped or underdoped samples the above mentioned
universality of the experimental results does not hold anymore.
For instance C. Kendziora et al.\cite{Kend96} reported for
overdoped Tl$_2$Ba$_2$CuO$_{6+\delta}$ (Tl-2201) a similar peak
position for the different symmetry components of the electronic
Raman scattering. The authors pointed out that the gap does not
scale with T$_c$, but rather decreases with an increase of doping,
yielding a $2\Delta_0/k_BT_c=3.9$. This led them to suggest that
in the overdoped Tl-2201 the order parameter has s-symmetry. One
should note, however, that existing calculations of the ELRS peak
positions (especially for the A$_{1g}$ scattering component
\cite{Dev95,Dev97,Car96,Carb96,Weng97}) strongly depend on the
chosen electronic structure and gap function parameters. For the
optimally doped Tl-2201 the difference between the peak positions
of the B$_{1g}$ and B$_{2g}$ components is about 10$\%$ only
\cite{Gasp97}. One can estimate an expected difference between the
corresponding peak positions for strongly overdoped Tl-2201 by
scaling the peak position of the B$_{1g}$ scattering component in
optimally doped Tl-2201 ($\approx\mbox{430~cm}^{-1}$) to that
reported for the strongly overdoped Tl-2201
($\approx\mbox{80-100~cm}^{-1}$). Such an estimate gives for the
strongly overdoped crystal a peak position of the B$_{2g}$
scattering component at only 8-10~cm$^{-1}$ lower frequency than
that of the B$_{1g}$ component. This is actually within
experimental error. Therefore the same position of the peaks
cannot prove s-wave pairing.

According to Devereaux et al.\cite{Dev95}, the low frequency power-law
behavior of the ELRS intensity is more "robust" concerning changes of
the FS topology as a result of overdoping and underdoping. Particularly the
$\omega^3$-law for the low frequency scattering in the B$_{1g}$ scattering
component and $\omega$-law for the A$_{1g}$ and B$_{2g}$ scattering
components should not change with doping in a d-wave superconductor.
Unfortunately the ELRS peaks in strongly overdoped Tl-2201 have their maxima
at a rather low frequency, which makes it difficult to determine their
low-frequency tails precisely. Additionally the low frequency scattering for
the A$_{1g}$ component is easily obscured by Rayleigh scattering. In order
to test the low frequency behavior in the overdoped Tl-2201 it is therefore
necessary to investigate moderately overdoped samples with a pair-breaking
peak not at too low frequency.

In addition to the scattering in the superconducting state the normal state
scattering provides important information about carrier dynamics.
Raman scattering in the normal state in channel L and assuming a single
impurity scattering lifetime $\tau$ can be described by a Lorentzian:

\begin{equation}\label{norm}
\Im\mbox{m}\chi_L(\omega, T>T_c)=2N_F{\gamma^2_L}\frac{\omega\tau}
{(\omega\tau)^2+1},
\end{equation}

where $\Gamma=1/\tau$ is the scattering rate, $\gamma_L$ is a
Raman vertex, and N$_F$ is the carrier density of states at the
Fermi level\cite{Zav90,Kost92}. Generally speaking, $\tau$ is a
function of the scattering channel $L$ and momentum
$\vec{k}$\cite{Mis94}. $\Im\mbox{m}\chi_L(\omega, T>T_c)$ has a
peak at the frequency $\omega=1/\tau$, and the spectrum falls off
as $1/\omega$. Using this fact one can analyze Raman spectra in
the normal state and determine how scattering rates change with
doping. Hackl et al.\cite{Hacl96} fitted their data for Bi-2212
using Eq.\ref{norm} and a frequency dependence of $\Gamma$ given
by the nested Fermi liquid model\cite{Viros92}. The scattering
rates at T$\approx$ 100~K were found to be $\Gamma(B_{1g})\approx$
600~cm$^{-1}$, $\Gamma(B_{2g})\approx$ 170~cm$^{-1}$ for the
nearly optimally doped Bi-2212 and $\Gamma(B_{1g})\approx$
160~cm$^{-1}$, $\Gamma(B_{2g})\approx$ 120~cm$^{-1}$ for overdoped
Bi-2212\cite{Hacl96}.

In this paper we present  electronic Raman scattering experiments
on moderately overdoped Tl$_2$Ba$_2$CuO$_{6+\delta}$ with
T$_c$=56~K. These are compared with measurements on optimally
doped (T$_c$=80~K) and strongly overdoped (T$_c$=30~K) crystals.
We show that similarly to optimally doped Tl-2201 also  moderately
overdoped Tl-2201 samples show a $\omega^3$-low frequency behavior
of the B$_{1g}$ scattering component and a linear low frequency
behavior for the B$_{2g}$ scattering component. The above
mentioned power laws are consistent with d-wave symmetry of the
order parameter. Additionally we will  discuss the changes of the
relative intensities of the pair breaking peaks in the
 A$_{1g}$ and B$_{1g}$ scattering components with doping, as well as the
electronic Raman scattering in the normal state.

\section*{Experimental}

We investigated the electronic Raman scattering in the
single-CuO$_2$ layered compound Tl-2201. This provides a
single-sheeted Fermi surface \cite{Tat93}. Therefore the
inter-valley scattering due to the multi-sheeted Fermi
surface\cite{Car96} invoked for the explanation of unexpectedly
large A$_{1g}$ scattering intensity does not play a role. Our
samples had a shape of rectangular platelets with the size of
2x2x0.15mm. Moderately overdoped and strongly overdoped crystals
of Tl-2201 were characterized by a SQUID magnetometer, T$_c$ was
found equal to 56$\pm$2~K (moderately overdoped) and 30$\pm$2~K
(strongly overdoped), respectively. The orientation of the
crystals was controlled by X-ray diffraction. The Raman
measurements were performed in quasi-backscattering geometry.
Raman scattering was excited using an Ar$^+$-ion laser. The laser
beam with 3mW power was focused into a spot of 50$\mu$m diameter.
The laser induced heating was estimated by increasing the laser
power level at a fixed temperature (5~K) and comparing the
dependence of the ELRS B$_{1g}$-peak intensity on laser power with
the temperature dependence of the intensity of this peak measured
at fixed laser power (3mW). Estimated additional heating was found
to be about 12.5$\pm$2.5~K (all data are plotted with respect to
the estimated temperature). In order to analyze pure scattering
geometries we extracted the A$_{1g}$ scattering component from the
X'X' (A$_{1g}$+B$_{2g}$) and XY (B$_{2g}$) scattering geometries.
The X' and Y' axes are rotated by 45$^{\circ}$ with respect to the
X and Y-axes. The X- and Y-axes are parallel to the Cu-O bonds in
the CuO$_2$ plane of the Tl-2201 unit cell. After subtraction of
the dark counts of the detector the spectra were corrected for the
Bose-factor in order to obtain the imaginary part of the Raman
response function. In order to analyze the low frequency behavior
of the B$_{1g}$ scattering component in moderately overdoped
Tl-2201 with T$_c$=56~K we performed measurement in superfluid He
(T=1.8~K). This gives us several advantages: Because of the huge
thermal conductivity of superfluid helium we do not have any
overheating of the sample due to laser radiation. The absence of
overheating allows us to precisely determine the real temperature
of the excited volume. For T=1.8~K the Bose factor is equal to
zero down to at least 10~cm$^{-1}$. Therefore down to 10~cm$^{-1}$
we actually measure the imaginary part of the Raman response
function.

\section*{Results and discussion}

The Raman spectrum of Tl-2201 shows several phonons and a broad
electronic continuum. The superconducting transition leads to the
redistribution of the continuum into a broad peak. In Figs.2-4 we
show the  B$_{1g}$, A$_{1g}$ and B$_{2g}$ scattering components of
the Raman scattering for $T \ll T_c$ (solid line) and $T > T_c$
(dashed line) for the Tl-2201 single crystals with T$_c$ =80~K
(Fig.2), T$_c$ =56~K (Fig.3), and T$_c$ =30~K (Fig.4). In order to
emphasize the redistribution of the scattering intensity in the
superconducting state compared to the normal state we draw not
only the Bose-factor-corrected raw spectra (Figs.2, 3 and 4, upper
panel), but we subtract the spectra above T$_c$ from the spectra
well below T$_c$ (Fig.2, 3 and 4, lower panel). The positions of
the ELRS peaks in the superconducting state for different
scattering components as a function  of doping are summarized in
Table I.

It is generally accepted that the B$_{1g}$ scattering component reflects
much of the properties of the superconducting density of
states\cite{Carb96}. Therefore it is reasonable to analyze intensities of
other components relative to the B$_{1g}$ scattering component.

There are  several differences between optimally- and overdoped
crystals.

i) If one identifies the  peak in the B$_{1g}$ ELRS component as a
2$\Delta_0$ one obtains the reduced gap value 2$\Delta_0/k_BT_c
\approx 7.8$ for the optimally doped crystal, while in the
overdoped crystals 2$\Delta_0/k_BT_c$ is close to 3, (see Table
I).

ii) For the optimally doped crystals the peak positions of the
B$_{2g}$ and A$_{1g}$ scattering components are lower than that of
the B$_{1g}$, (see Fig.2 and Table I). In the overdoped crystals
the B$_{2g}$ component peaks at a frequency very close to that of
the B$_{1g}$ scattering component (see Figs.3, 4 and Table.I.),
although its peak position is still about 10$\pm$2\% lower
(similar to the optimally doped Tl-2201, see Table I). The
A$_{1g}$ peak position is close to that of the B$_{1g}$ peak as
well, although an exact determination of the pair-breaking peak
position for the A$_{1g}$ scattering component is difficult due to
the A$_{1g}$ phonon at 127~cm$^{-1}$ of moderately overdoped
Tl-2201, (see Fig.3) or due to the superimposed Rayleigh
scattering in strongly overdoped Tl-2201
 (see Fig.4).

iii) The most drastic changes of the relative ELRS peak intensity with
doping are seen in the A$_{1g}$ scattering component. For the optimally
doped crystal we observe a strong peak, which is comparable in intensity to
that of the B$_{1g}$ component, see Fig.2a and b, lower panel. In contrast,
for two  overdoped crystals (Figs.3, 4 a and b, lower panel) the relative
intensity of the ELRS peak in the A$_{1g}$ scattering component is weak.

iiii) In contrast to the A$_{1g}$ scattering component the
intensity of the B$_{1g}$ scattering component is stronger in the
moderately overdoped sample (Fig. 3a) compared to the optimally
doped one (Fig. 2a). For the strongly overdoped sample an exact
determination of the relative intensity of the pair-breaking peak
is difficult  in all scattering components. The pair-breaking peak
is at too low frequency ($\approx 60~cm^{-1}$), therefore its
intensity is very sensitive to the Bose-factor correction, which
in turn depends upon the uncertainty in the estimated temperature.
Additionally, Rayleigh scattering and impurity induced
scattering\cite{Dev95} may obscure the evaluated difference
between the corresponding spectra below and above T$_c$.

According to Devereaux et al.\cite{Dev95}, the $\omega^3$-law for
the low frequency scattering in the B$_{1g}$ scattering component
and the $\omega$-law for the A$_{1g}$ and B$_{2g}$ scattering
components should not change with doping in d-wave
superconductors. In order to check these power laws for the
moderately overdoped Tl-2201 we have performed measurements in
superfluid helium (T=1.8~K). To illustrate the low frequency
behavior of the imaginary part of the Raman response function in
the B$_{1g}$ and B$_{2g}$ scattering components on the same
frequency scale we have scaled the Raman shift by the
corresponding peak position, as shown in Fig. 5a. The fit of the
low frequency scattering in the B$_{1g}$ scattering component with
the $\omega^n$-function leads to exponents n=2.9 and 3.5 for the
optimum doped and moderately overdoped Tl-2201, respectively. An
even better fit to the low frequency scattering intensity in
moderately overdoped Tl-2201 was obtained with a linear term added
to the $\omega^n$ function, similarly to overdoped
Bi-2212\cite{Hacl96}. The appearance of such a crossover from
linear to a power law in the B$_{1g}$ scattering component
indicates the presence of impurities\cite{Dev95}. For the B$_{2g}$
scattering component one can easily fit the low frequency
scattering of optimally to overdoped samples with a
linear-in-$\omega$ law as shown in Fig.5b. Unfortunately in the
T$_c=$30-K crystal the expected ELRS peak is too close to zero
frequency to make a definite conclusion about its low frequency
behavior. The observed power laws ( Fig.5) lead to the conclusion
that even overdoped Tl-2201 has a d-wave symmetry of the order
parameter.

Let us now discuss temperature induced spectral changes in the
overdoped crystal. A detailed temperature dependence for the
Tl-2201 (T$_c$=56~K) sample is shown for the B$_{1g}$ component in
Fig.6. With increasing temperature the intensity of the
pair-breaking peak decreases and its position shifts toward lower
frequency. This dependence slightly differs from that predicted by
the BCS theory, as shown in the inset of  Fig.6, i.e. the gap
opens more abruptly. At the same time the intensity of the
pair-breaking peak decreases nearly linearly with increasing
temperature (see insert in Fig.7) whereas the intensity of the low
frequency scattering (at for instance $\approx$50~cm$^{-1}$)
increases. At a temperature close to T$_c$ both intensities match.
From this data one can determine the ratio of the superconducting
response to the normal state response in the static limit ("static
ratio"), i.e when $\omega\rightarrow 0$ and compare it with the
calculations of the ratio in the presence of
impurities\cite{Dev97}. From such a comparison we found for the
moderately overdoped Tl-2201 the corresponding value of the
scattering rate to be $\Gamma/\Delta(0)\approx0.5$. This leads to
$\Gamma\approx$60~cm$^{-1}$. In the normal state spectra (we
discuss the imaginary part of the Raman response function) one
sees an increase of the intensity towards zero with a broad peak
at $\approx$50~cm$^{-1}$, Figs.3 and 7. This peak is more
pronounced in the B$_{1g}$ scattering component. Such a peak can
be attributed to impurity induced scattering. According to
Eq.\ref{norm} the frequency of the peak corresponds to the
scattering rate $\Gamma=1/\tau$ of the normal
state\cite{Zav90,Kost92}. The position of the peak depends
strongly on doping. It is roughly 35 or 50~cm$^{-1}$ for strongly
and moderately overdoped Tl-2201, respectively. Practically there
is no anisotropy of the peak position comparing the B$_{1g}$ and
B$_{2g}$ scattering components. Note that the scattering rates
calculated from the peak positions are very close to that
evaluated from the "static ratio" and sufficiently smaller than
that found by Hackl et al.\cite{Hacl96} using a frequency
dependence of $\Gamma$ given by the nested Fermi liquid model.
Scattering rates may also be determined using the frequency
dependent conductivity from the infrared measurements. One finds
for many HTSC scattering rates $1/\tau$ of about
100-200~cm$^{-1}$ at T$\approx$100~K\cite{Tim92}. Additionally and
very surprisingly, the scattering rates decrease with increasing
overdoping\cite{Tim96}. From our Raman measurements we found
scattering rates $\Gamma=1/\tau$=35 or 50~cm$^{-1}$ for strongly
and moderately overdoped Tl-2201 not too far from the infrared
data, and a similar decrease of $\Gamma$ with increasing
overdoping.

We would like to sum up the effects of overdoping that are also partly
observed in other HTSC:

In the nearly optimally doped regime
(YBCO\cite{Hackl88,Coop88,Chen93}, Bi-2212\cite{Stauf92},
Tl-2223\cite{Hof94}, La-214\cite{Chen94},
Tl-2201\cite{Nem93,Gasp97}) the ELRS peak positions scale with
T$_c$ for all scattering components. The B$_{1g}$ scattering
component is most sensitive to changes of T$_c$. The relative
intensity of the ELRS A$_{1g}$ peak is stronger or at least
comparable to that of the B$_{1g}$ component. The relative
intensity of the B$_{2g}$ peak is always the weakest one.

For the overdoped crystals (Tl-2201, Bi-2212)\cite{Kend96,Hacl96}
the peak position of the B$_{1g}$ scattering component decreases
faster than T$_c$ so that 2$\Delta_0/k_BT_c$ decreases with
overdoping from 7.4 to $\approx$3 (data of this paper), or from 8
to 5 in Bi-2212\cite{Hacl96}. The relative intensity of the
A$_{1g}$ ELRS peak as compared to B$_{1g}$ decreases when the
system becomes more overdoped\cite{Blum97}. This is an important
point concerning the influence of the Fermi surface topology
changes on Raman scattering and will be discussed further below.

We will now discuss some reasons which may explain the shift of the
B$_{1g}$ peak position with doping. The decrease of the B$_{1g}$ ELRS peak
position and 2$\Delta_0/k_BT_c$ with doping is connected to the fact that
the crossing of the Fermi surface with the Brillouin zone moves away from
the (0,$\pm\pi$), ($\pm\pi$,0) points with doping. Therefore the FS average
$\langle\gamma^2_{\vec{k}}\lambda_{\vec{k}}\rangle$ of the Raman vertex
with the Tsuneto function in Eq.\ref{rares} gives a $\Delta_0$ smaller than
$\Delta_{max}$. A detailed discussion of this poin is given in the
work of Branch and Carbotte \cite{Carb96}. In the case of optimum doping
it is supposed that the Fermi level is close to the van Hove singularity
(vHs) so that the FS pinches at the (0,$\pm\pi$), ($\pm\pi$,0) points of
the BZ\cite{Nov95} leading to  $\Delta_0\approx\Delta_{max}$.

Now let us turn to the decrease of the A$_{1g}$ vs. B$_{1g}$
intensities of the ELRS with doping. In contrast to B$_{1g}$ and
B$_{2g}$ the A$_{1g}$ scattering component is affected by the
screening term. We suppose that "screening" itself is connected
with the FS anisotropy, which is in turn affected by the van Hove
singularity. In optimally doped crystals vHs is close to the Fermi
level (FL) leading to strongly anisotropic FS. By overdoping we
move FL from vHs. This  leads to a more isotropic FS with larger
"screening". Therefore the increase of "screening" with doping
would be a plausible explanation for the observed decrease of the
A$_{1g}$ scattering component with doping.

This suggestion has a consequence for the intensity of the
B$_{1g}$ scattering component. Namely the "screening" term for the
A$_{1g}$ scattering component has the same symmetry as the bare
term for the B$_{1g}$ scattering component (see Fig.1). If we
suppose that the "screening" increases, the B$_{1g}$ response
should also increase.  This is in  agreement with our results (see
Figs. 2a and 3a, lower panel).

In conclusion we have presented measurements of the electronic
Raman scattering on optimally doped as well as moderately and
strongly overdoped Tl-2201 single crystals. The strong decrease of
the A$_{1g}$  scattering intensity with increasing overdoping has
been observed. We connect this effect with the changes of the FS
topology connected to the existence of a van Hove singularity. We
propose investigations on other overdoped HTSC in order to check
this idea.  Our measurements of the low frequency behavior of the
electronic Raman scattering in optimally doped and moderately
overdoped Tl-2201 confirmed a d-wave symmetry of the order
parameter, in contrast to earlier reports \cite{Kend96}. The
scattering rates, we have evaluated from the normal state Raman
spectra as well as a decrease of them with overdoping are
consistent with the existing infrared data.

\section*{Acknowledgments}
This work was supported by DFG through SFB 341, BMBF FKZ 13N7329
and INTAS grants 94-3562 and 96-410. One of us (L.V.G.)
acknowledges support from the Alexander von Humboldt Foundation.

\begin{figure}
\caption{Schematic representation of the Raman response function
in Eq.\ref{rares} due to the breaking of Cooper pairs and
including Coulomb repulsion. The  Tsuneto function $\lambda_{\vec{k}}$
is represented as a squared gap function
$\Delta^2(\vec{k})=\Delta^2_{max}\cos^22\phi$. Raman vertices are
chosen as $\gamma_{B_{1g}} \propto \cos2\phi$,
$\gamma_{B_{2g}}\propto\sin2\phi$, $\gamma_{A_{1g}} \propto
1+\cos4\phi$, where $\phi$ is an angle between $\vec{k}$ and the Cu-O
bond direction within the CuO$_2$ plane.}
\end{figure}

\begin{figure}
\caption{Imaginary part of the Raman response in optimally doped
Tl-22O1 (T$_c$=80$\pm$5~K) at T=20~K (solid curve), and T=110~K
(dashed curve) for the a) B$_{1g}$, b) A$_{1g}$, and c) B$_{2g}$
scattering components (upper panel) and for the corresponding
subtracted spectra:
$\Im\mbox{m}\chi$(T=20~K)$-\Im\mbox{m}\chi$(T=110~K)(lower
panel).}
\end{figure}

\begin{figure}
\caption{Imaginary part of the Raman response in moderately
overdoped Tl-2201 (T$_c$=56$\pm$2~K) at T=20~K (solid curve), and
T=75~K (dashed curve) for the a) B$_{1g}$, b) A$_{1g}$, and c)
B$_{2g}$ scattering components (upper panel) and for the
corresponding subtracted spectra:
$\Im\mbox{m}\chi$(T=20~K)$-\Im\mbox{m}\chi$(T=75~K)
 (lower panel).}
 \end{figure}

\begin{figure}
\caption{Imaginary part of the Raman response in strongly
overdoped Tl-2201 (T$_c$=30$\pm$2~K) at T=15~K (solid curve), and
T=50~K (dashed curve) for the a) B$_{1g}$, b) A$_{1g}$, and c)
B$_{2g}$ scattering components (upper panel), and for the
corresponding subtracted spectra: $\Im\mbox{m}\chi$(T=15~K)$-\Im\mbox{m}\chi$(T=50~K) (lower panel).}
\end{figure}

\begin{figure}
\caption{Imaginary part of the Raman response in optimally doped
(T$_c$=80$\pm$5~K, dashed line, at T=20~K), moderately overdoped
(T$_c$=56$\pm2$~K, dash-dotted line, at T=1.8~K), and strongly
overdoped (T$_c$=30$\pm$5~K, dotted line, at T=15~K) Tl-2201 for
the a) B$_{1g}$ and b) B$_{2g}$ scattering components. For each
doping and scattering component the frequency-axis is rescaled to
the position of the respective pair-breaking peak. The solid
curves show fits to the low frequency scattering with the a)
$\Im\mbox{m}\chi\sim \omega^n$ (n=2.9 and 3.5 for the crystals
with T$_c$=80$\pm$5~K and 56$\pm$2~K, respectively) and b)
$\Im\mbox{m}\chi\sim \omega$ function.}
\end{figure}

\begin{figure}
\caption {Temperature dependence of the Raman scattering intensity
of the B$_{1g}$ scattering component in the moderately overdoped
Tl-2201 (T$_c$=56$\pm$2~K) without corrections of the Bose-factor.
With increase of the temperature the pair-breaking peak position
of the ELRS shifts to lower frequency and its intensity decreases.
The insert shows the temperature dependence of the pair-breaking
peak position (solid circles) and the expected dependence from
BCS theory (solid line). Note that at T=59~K$>$T$_c=56~K$ one sees
a characteristic increase of the intensity towards zero frequency
which is attributed to impurity induced scattering.}
\end{figure}

\begin{figure}
\caption {Imaginary part of the Raman response function at
T$\le$T$_c$, with T=55~K (solid line), T=48~K (dashed line) and
T=45~K (dotted line) for the B$_{1g}$ scattering component of
moderately overdoped Tl-2201. Arrows show the pair-breaking peak
($\approx$ 105~cm$^{-1}$ at T=45~K) and the peak due to scattering
on the "normal excitations" ($\approx$ 50~cm$^{-1}$). The insert
shows a temperature dependence of the pair-breaking peak intensity
(solid squares) and the intensity of the "normal excitation"
scattering (open squares) in the B$_{1g}$ scattering component in
the moderately overdoped Tl-2201 (T$_c$=56$\pm$2~K).}
\end{figure}

\begin{table}
\caption{Peak positions of the B$_{1g}$ , A$_{1g}$  and  B$_{2g}$
electronic Raman scattering components, for optimally and overdoped
Tl-2201.}
\begin{tabular}{cccccc}
Compound Tl-2201 &  T$_c$ [K] & B$_{1g}$ [cm$^{-1}$] &
A$_{1g}$ [cm$^{-1}$]& B$_{2g}$ [cm$^{-1}$] & 2$\Delta_{max}/k_BT_c$\\
\tableline
optimally doped & 80$\pm$5 & 430$\pm$15 & 345$\pm$20 & 380$\pm$35 &
7.8$\pm$0.4\\
moderately overdoped & 56$\pm$2 & 125$\pm$10 & 110$\pm$20 & 120$\pm$10 &
3.3$\pm$0.3\\
strongly overdoped & 30$\pm$2 & 60$\pm$5 & ? & 50$\pm$5 & 2.9$\pm$0.4
\end{tabular}
Question mark (?) indicates  no detection of  a pair-breaking peak.
\end{table}

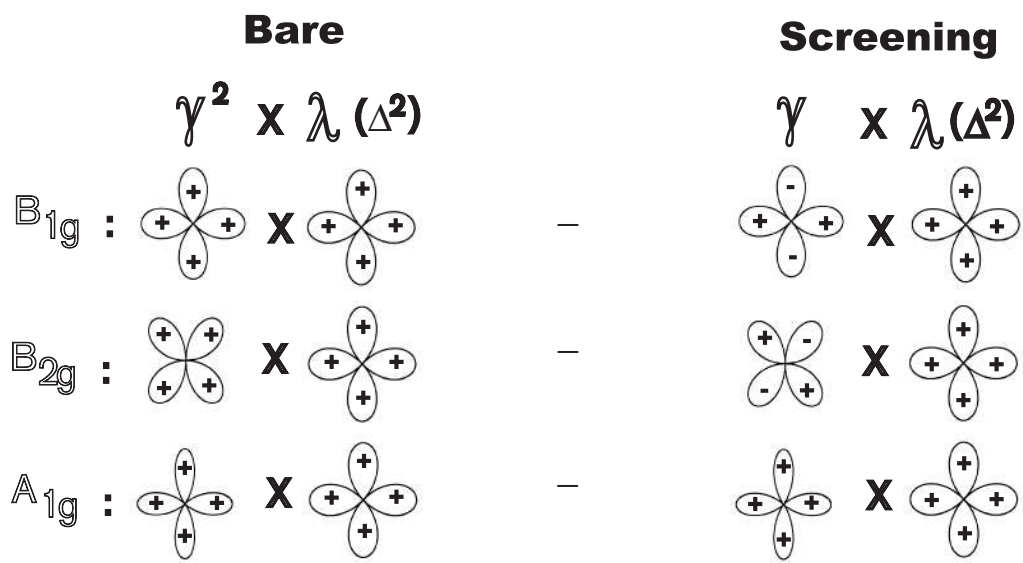
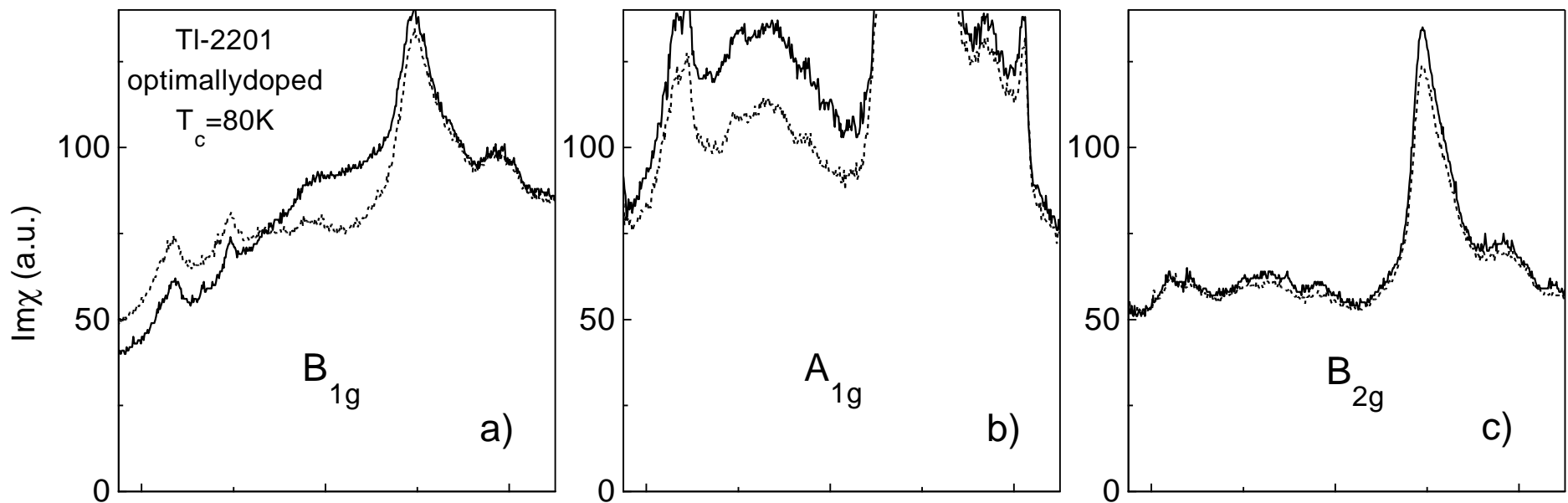
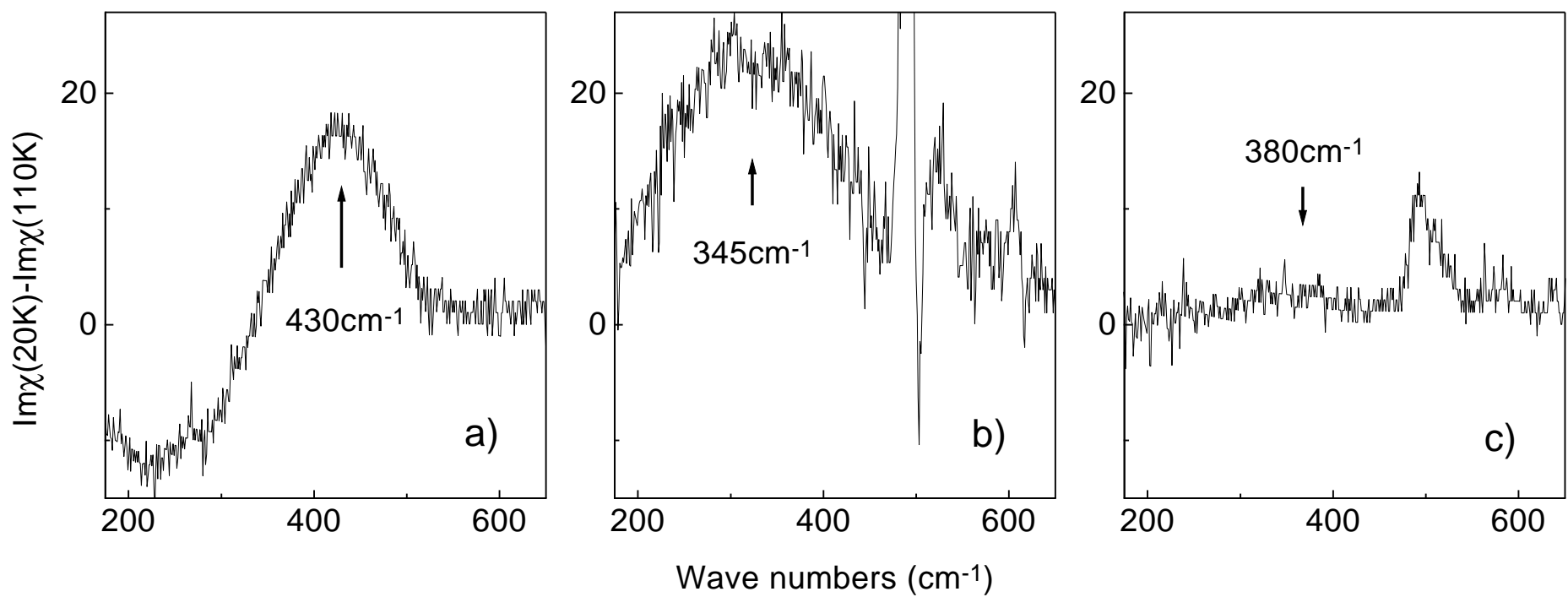
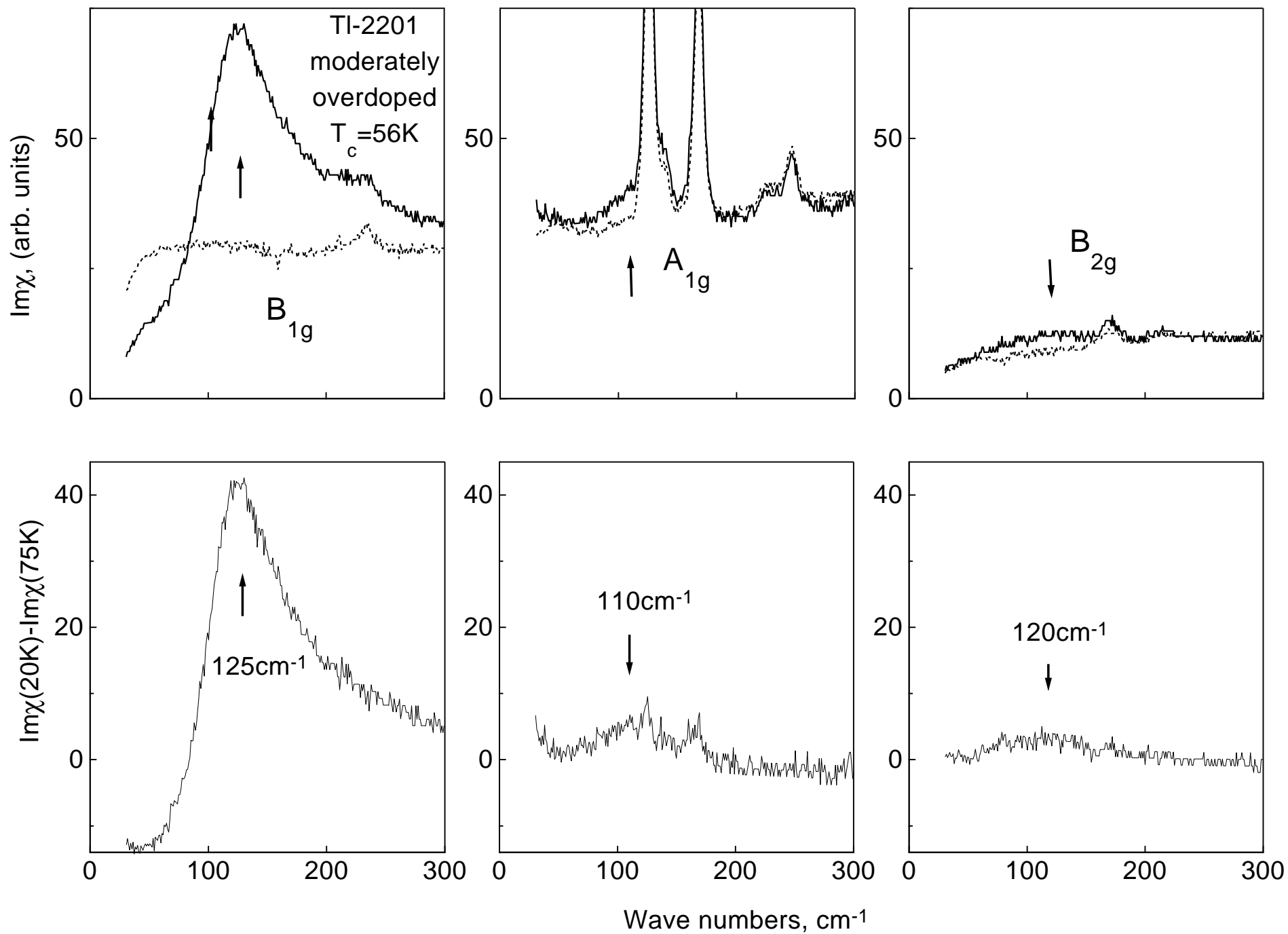
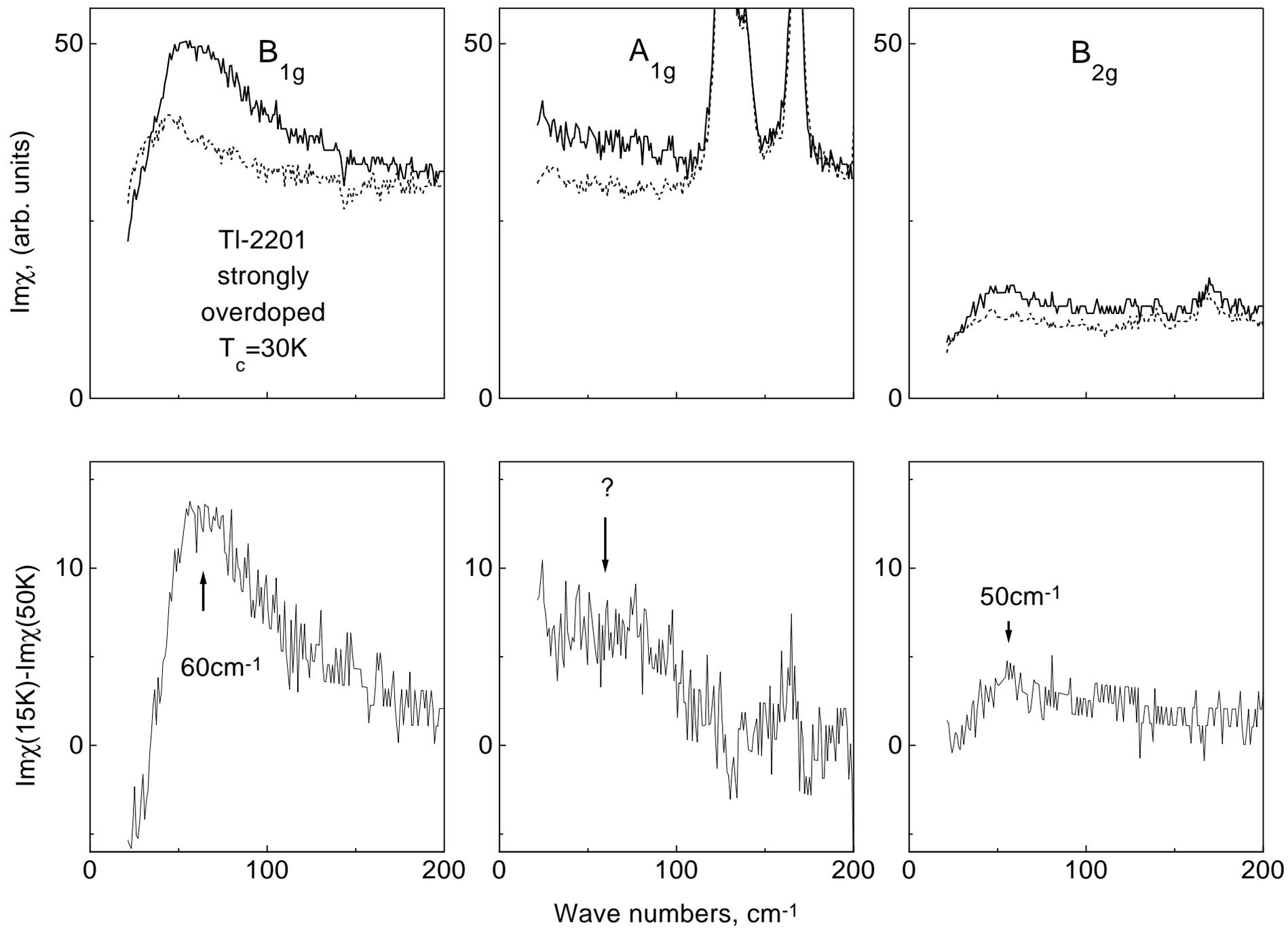
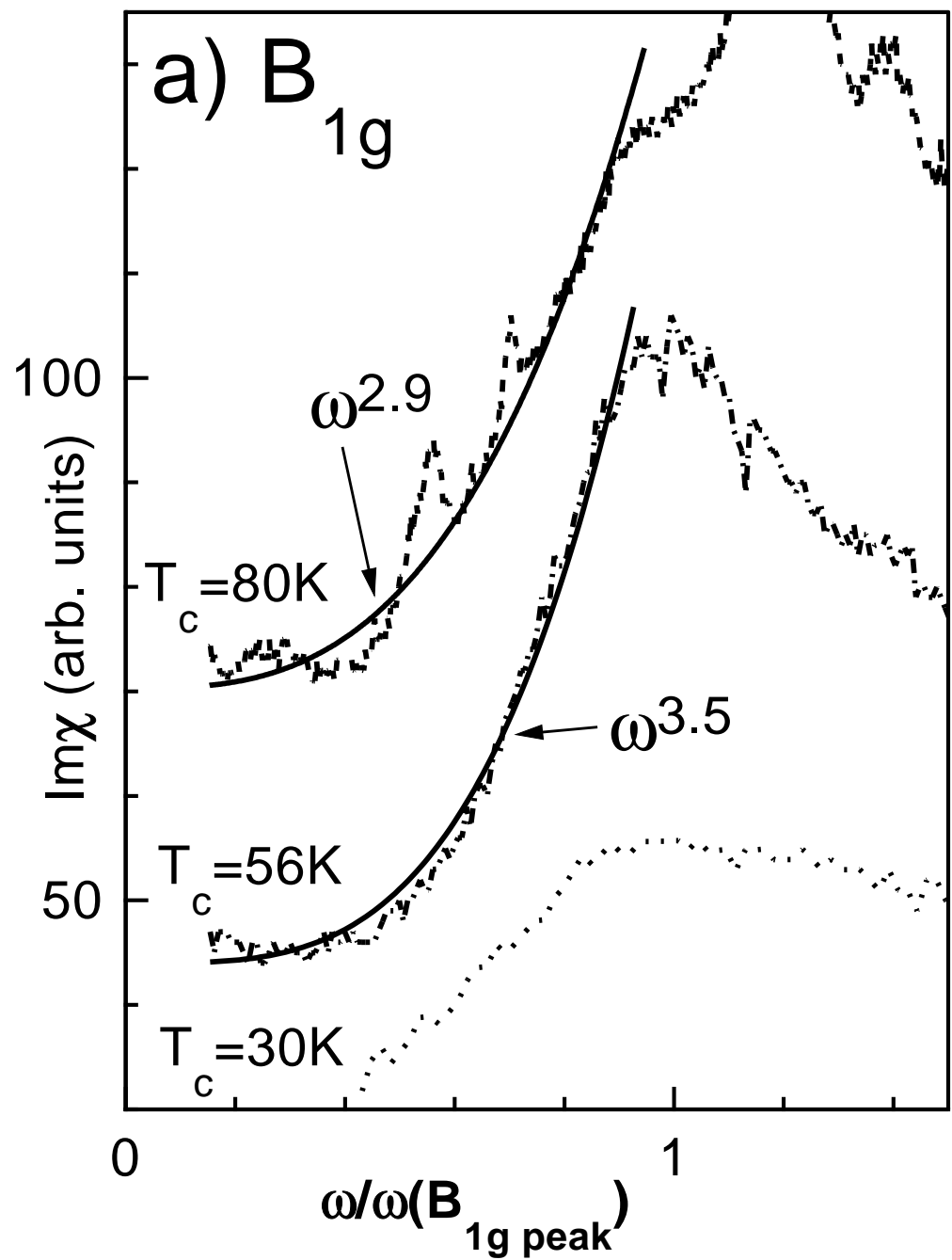
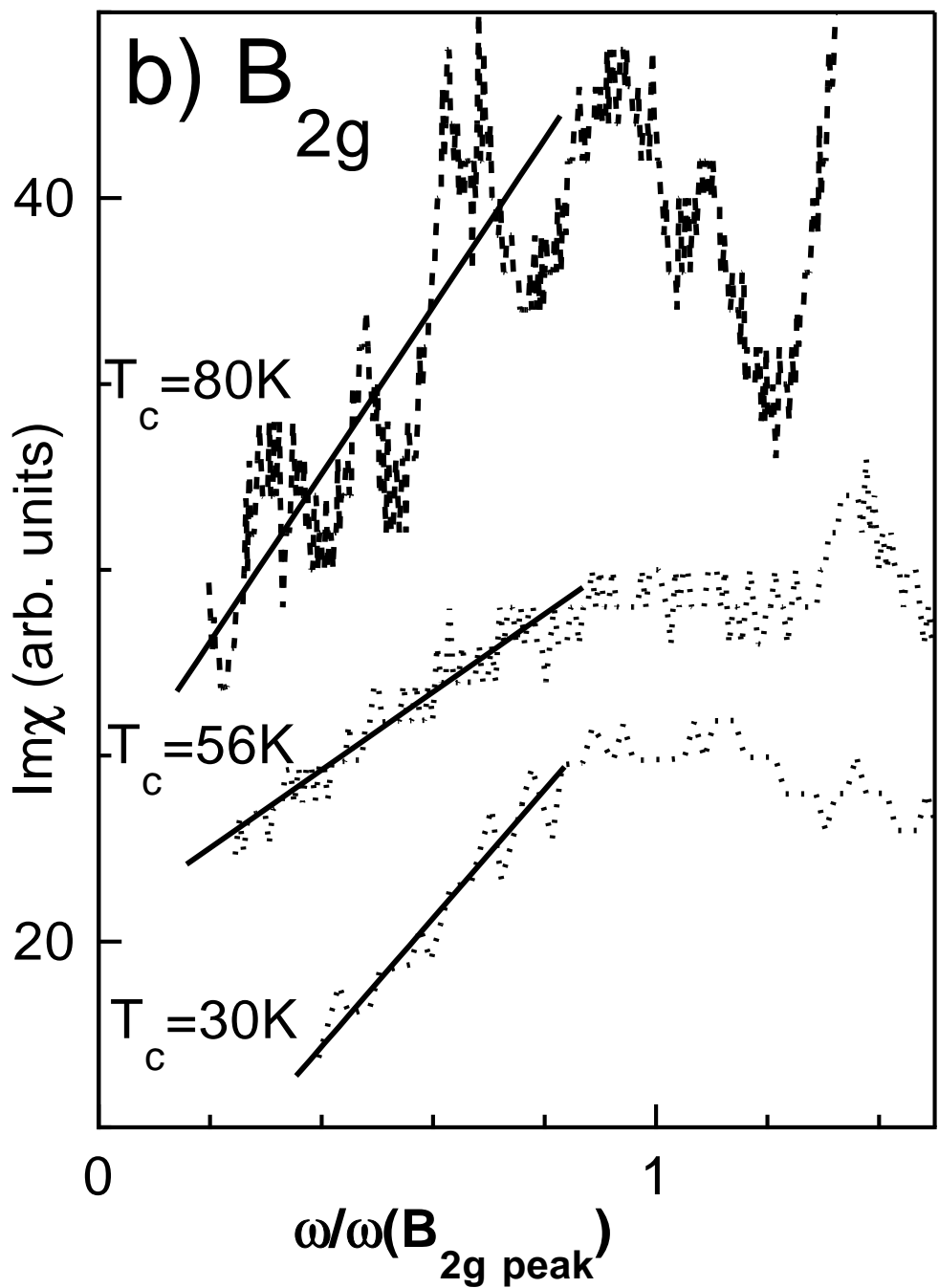
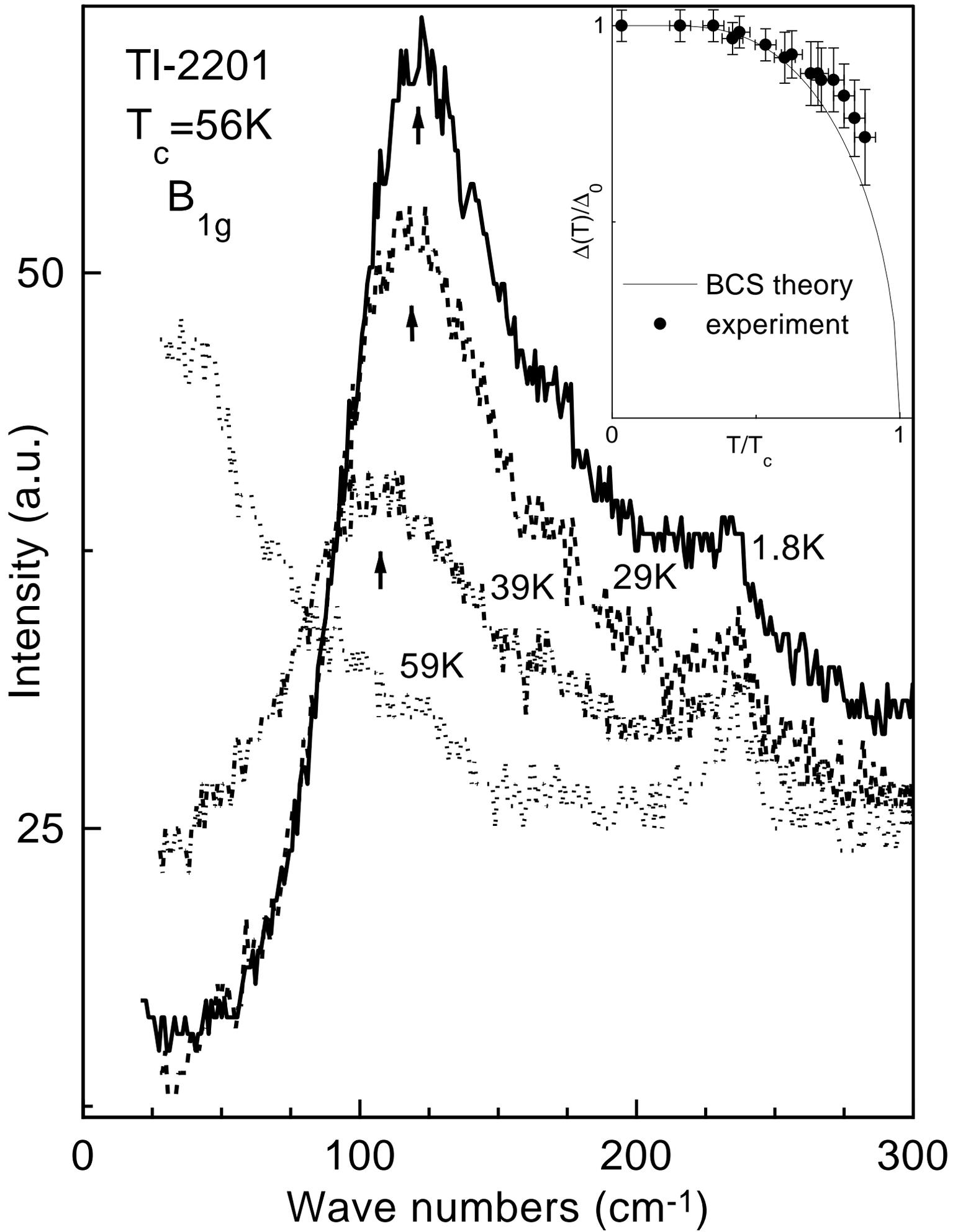
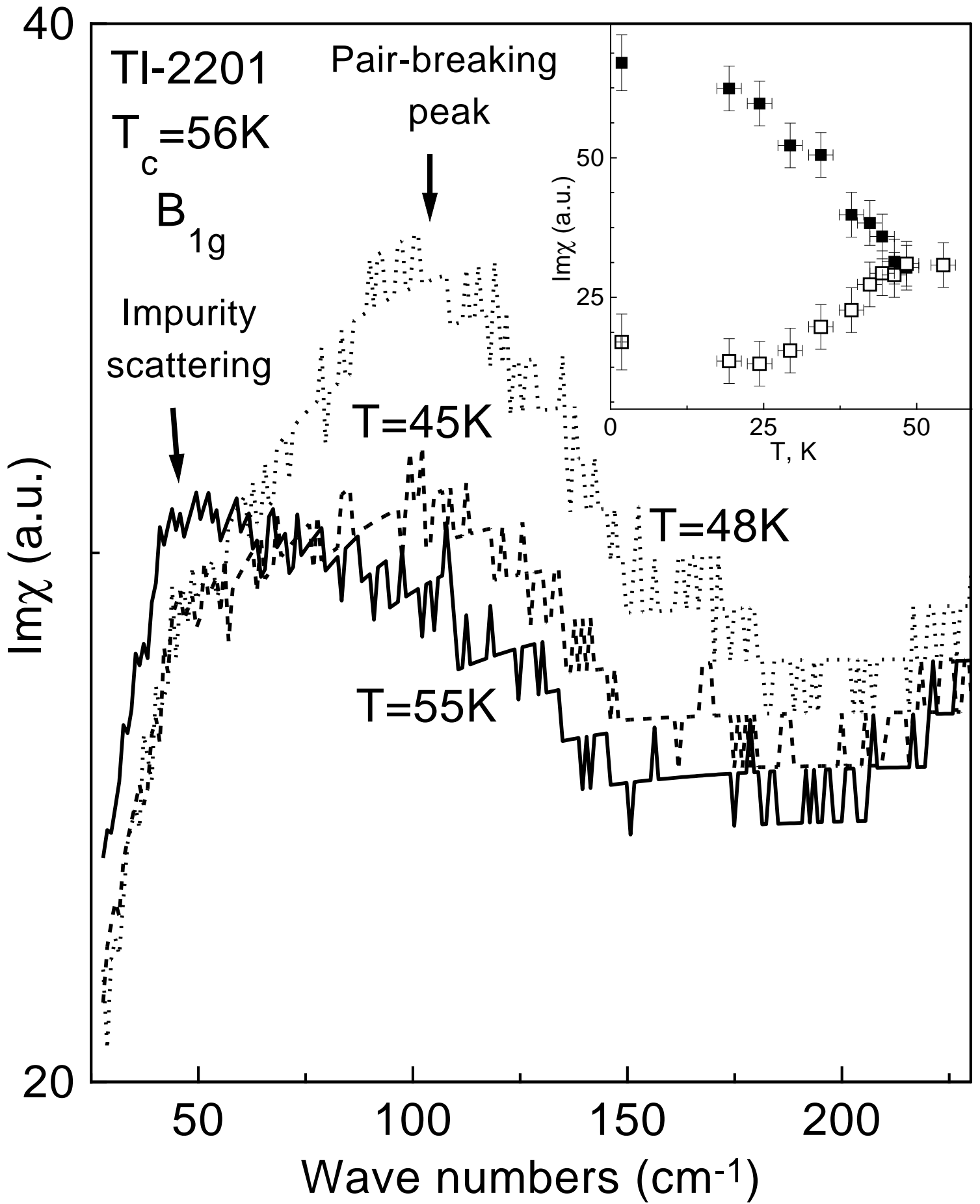
\end{document}